# Parallel and series conduction model in Topological Insulators

**Sourabh Singh[#], R K Gopal[#], Jit Sarkar, and C. Mitra**
*Indian Institute of Science Education and research Kolkata, Mohanpur- 741 246 India*

E-mail: *sourabh1220@iiserkol.ac.in*

**Abstract**. In the past few years there has been a surge in the material science engineering in order to synthesize bulk insulating and surface metallic Topological Insulating (TI) materials. This quest is not only theoretically important but also promising from the novel application perspective. The dependence of temperature on resistance (R-T) of a particular sample reveals a plethora of information about the electronic properties especially in a unique sample like TI where there are two components comprising of an insulating bulk and metallic surface states. Depending on the amount of intrinsic doping during the sample formation, the bulk can either couple or remain decoupled with the surface. The former leads to a metallic R-T profile whereas the latter is captured by an insulating R-T behavior. These two behaviors can be represented by series and parallel resistor models respectively. In this work we study the R-T behavior in the framework of resistor models capturing the essential features of our sample.

## Introduction

Topological insulators are a new state of quantum matter possessing unusual physical properties due to its linearly dispersed surface states in the bulk band gap [1]. These states are known as Dirac states owing to the linear dispersion branches spanning from bulk valence band to bulk conduction band [2]. This unique arrangement of branches in the bulk band gap makes them distinct from other type of trivial surface states such as Rashba split surface states [3]. The main mechanism behind the appearance of these states is bulk band inversion due to strong spin orbit coupling [4]. This class of materials possesses massless helical surface states which are protected against the non-magnetic disorder due to the presence of spin-orbit coupling. The helical nature of the surface states guaranties the absence of the backscattering from disorder as opposed to the trivial two dimensional electron gas systems which localize in the presence of disorder at low temperatures.

Previously discovered 3D TI material systems $Bi_2Se_3$ (BS), $Bi_2Te_3$ (BT) and $Sb_2Te_3$ (ST) were intrinsically doped during sample growth. This doping is due to the presence of Se vacancies or antisite defects resulting in a high carrier density in the bulk and shifting of Fermi level either in the bulk conduction band or valence band [5]. Thus the response of the surface Dirac states was masked by these residual bulk carriers. Despite showing clear signature of single Dirac cone structure in the ARPES and STM measurements, transport studies could not resolve surface dominated transport of such states from the bulk carriers. The temperature dependent resistivity of many of these materials were found to be metallic in nature [6]. The fate of a device response in most of the cases, in magnetotransport studies is inherited from the temperature dependent resistivity (R-T).

The two channels in the case of topological insulator are: Temperature independent surface channel and temperature dependent bulk channel. The R-T profile of topological insulators thus can be suitably mapped with parallel resistor model. We know that in a parallel circuit the electron will prefer the less resistive path for conduction. In the former case, when the system is coupled it is likely that the bulk overwhelms the surface conduction [7]. It is quite visible even in the R-T curve. The combined system behaves like a metal with the resistance coming down with the temperature analogous to a metal (see figure 2). In the second scenario the contribution of the bulk is feeble in comparison to the surface contribution. Therefore, it resembles a semiconducting activated kind of behaviour. This decoupled regime is a result of the competition between the metallic surface states and the insulating bulk. As the temperature is lowered the bulk contribution to the conductivity further quenches due to unavailability of thermal energy for the carriers to overcome the band gap.

## 2. Experimental Section

All the films were deposited on Si (100) substrates by Pulse Laser Deposition Technique (PLD). KrF excimer laser (wavelength= 248nm) was used to deposit these thin films. The deposited films were thoroughly characterized and then subjected to transport measurements. Thickness of the films were optimized by varying the no. of laser pulses. Insulating BST films were obtained by post deposition annealing at higher temperatures [8][9].

## 3. Results and Discussions

Reducing the temperature results in lowering of the resistance because of the contribution of temperature independent conductivity of the topological surface states. Depending on the relative amount of surface states to the bulk states this effect is vividly visible in the R-T profile.

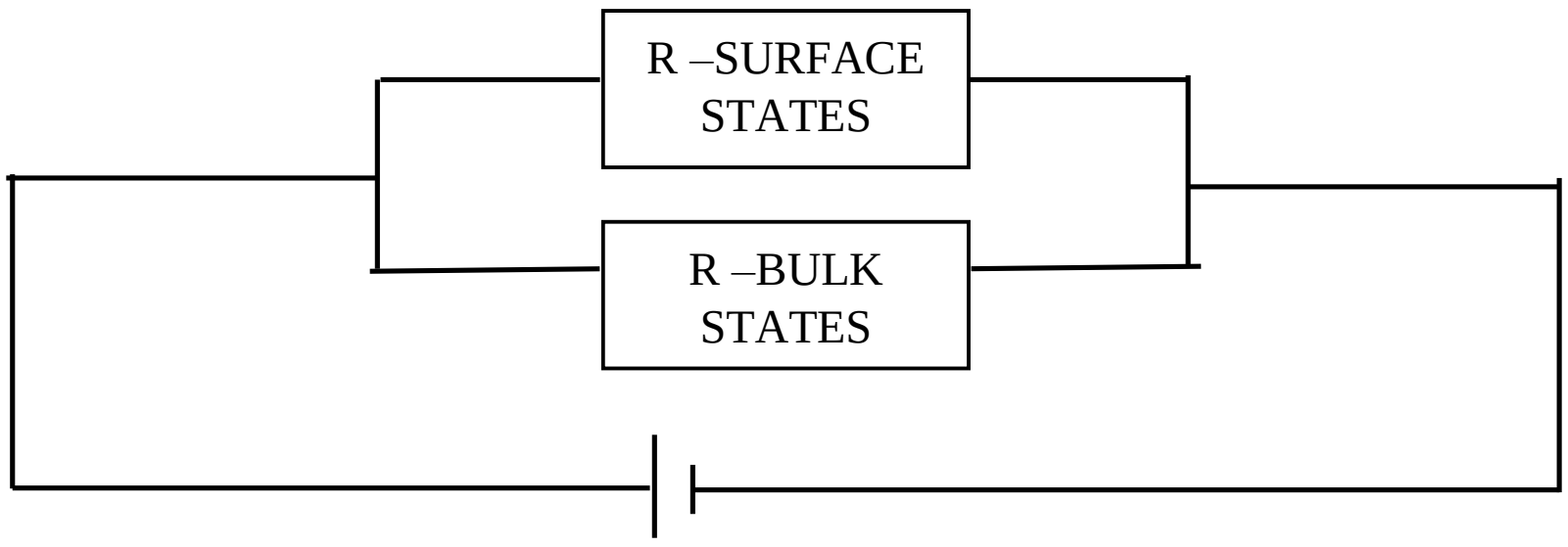

Figure.**1.** Schematic representation of parallel conduction channel in topological insulators.

The TI thus comprises of two types of conduction channels and they should be parallel to each other. But surface to bulk coupling plays an important role in establishing a relationship between the two independent channels. For a strongly coupled system the temperature dependent resistance can be fitted with a series model:

$$R\ (T) = R_1 + R_2$$

In the case of TI the above equation can be written as:

$$R(T) = R_0 + B^*\exp(-P/T) + C^*T^2 \qquad (1)$$

Where $R_0$ stands for resistance due to impurity scattering, P term corresponds to the phonon scattering and the last quadratic term is due to electron-electron interaction as obtained from Fermi liquid theory.

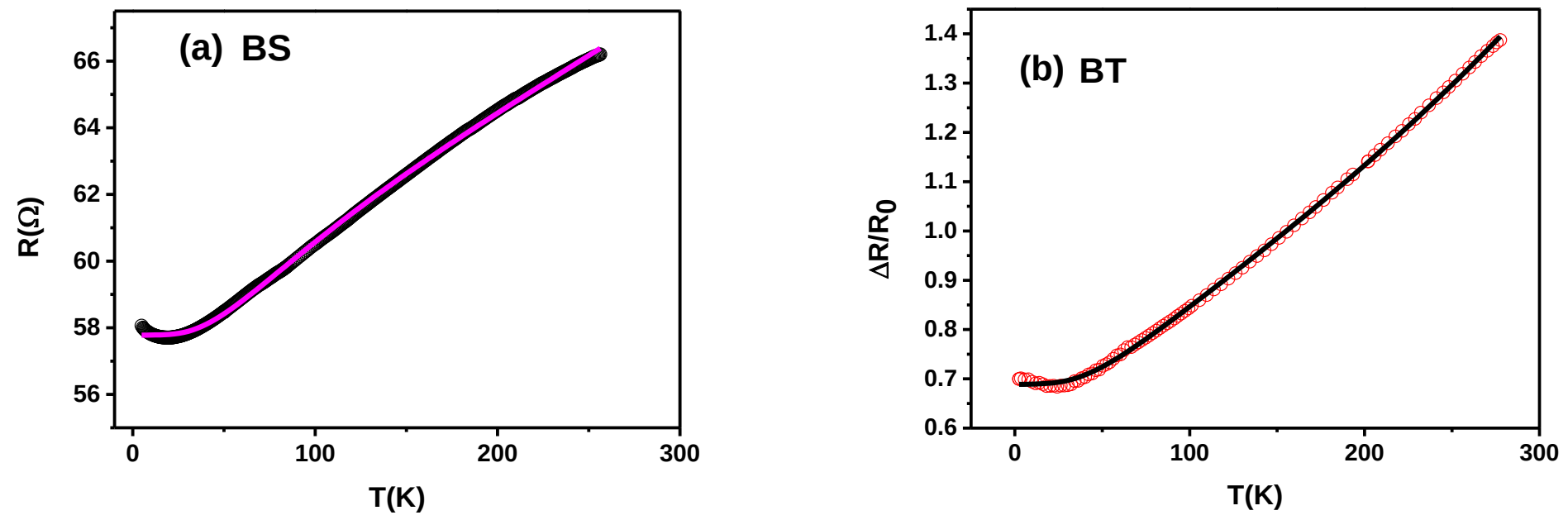

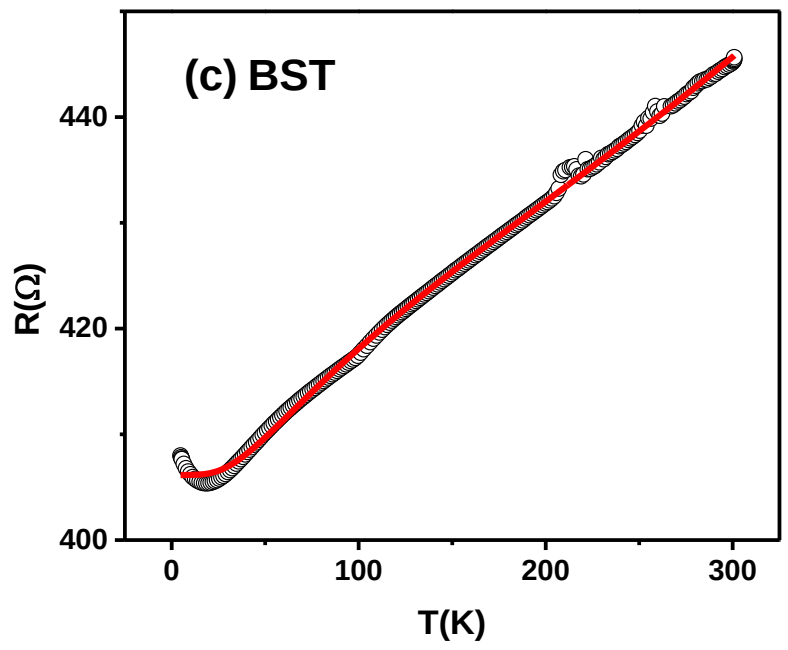

Figure.**2**. Resistance vs temperature profile of (a) BS (b) BT and (c) BST thin film. The data is fitted with a series resistive model given by equation (1)

From the fit the various parameters were obtained and a plethora of information regarding the sample characteristic is acquired from it. The coefficient C corresponds to electron-electron interaction strength. The value of C for BS, BT and BST are $4.07 \times 10^{-5}$, $5.61\times 10^{-6}$ and $1.95 \times 10^{-4}$ respectively. The other scenario is where there is minimum carrier density contribution from the bulk. So, there is no appreciable amount of surface to bulk coupling. For a weakly coupled system the R-T can be suitably described with a parallel resistor model.

$$\frac{1}{R} = \frac{1}{R_1} + \frac{1}{R_2}$$

The G-T profile in this case is fitted with a series conduction model given by the following equation:

$$\begin{aligned} G_t &= G_b + G_S \\ &= 1/\left(Re^{\Delta/T}\right) + 1/\left(A + BT\right) \end{aligned} \qquad (2)$$

$G_b(T) = 1/(Re^{\Delta/T})$ and $G_S(T) = 1/(A+BT)$, where Δ represents the activation energy for impurity states in the bulk. A and B represents the residual impurity scattering term and phonon scattering term respectively. The former term corresponds to the bulk contribution which dominates in the high temperature regime while the later term depicts surface state contribution which dominates the proceedings in the low temperature regime.

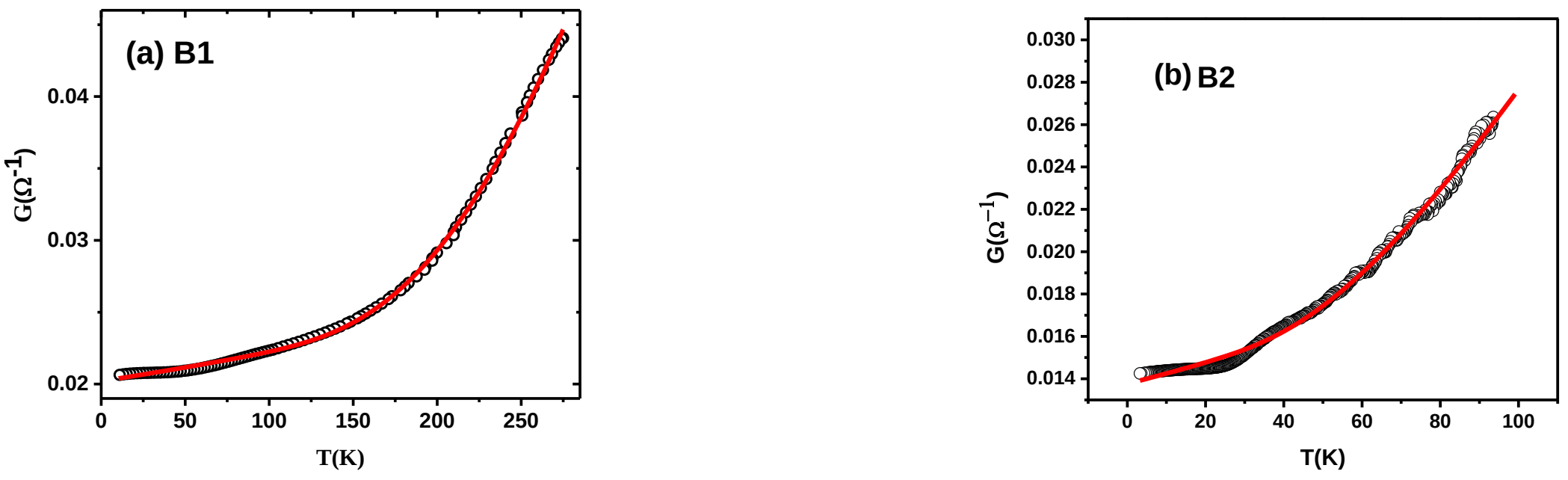

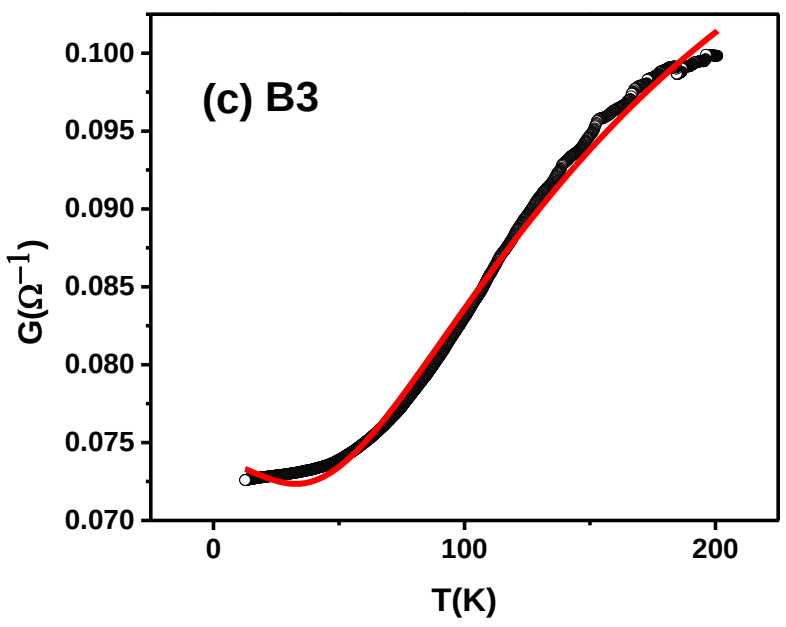

Figure.**3**. Resistance vs. temperature of three BST thin films of different thicknesses. (a) B1 corresponds to 500nm (b) B2 corresponds to 300 nm and (c) B3 corresponds to 150nm. The R-T curve is fitted with equation (2)

## 4. Conclusion

The bulk-boundary correspondence intrinsically connects the bulk and the surface states. The extent of surface to bulk coupling drastically changes the properties of TI materials. R-T curve depicts a new understanding in this regard as the resistance model varies from series to parallel. Furthermore, fitting the R-T curve with appropriate model yield parameters which explain the properties of the TI.

# Authors SS and RKG have contributed equally in this work.

**Acknowledgement**

The authors would like to thank Ministry of Human Resource Development, Govt. of India for financial assistance. SS and JS would like to thank University Grants Commission (UGC) for stipend and contingency grants. SS would like to thank ICPS 2016 organizers.